\documentclass[aps,prd,letterpaper,11pt,twoside,tightenlines,nofootinbib,showpacs,preprint,twocolumn]{revtex4}
\usepackage[T1]{fontenc}
\usepackage[utf8]{inputenc}
\usepackage[english]{babel}
\usepackage[sort&compress]{natbib}

\usepackage{graphicx}
\usepackage{epsfig}
\usepackage{graphicx}

\usepackage{amsmath}

\usepackage[figurename=FIG.,tablename=TABLE,font=footnotesize]{caption}

\begin{document}
%%%   New Definitions
\newcommand{\eg}{{\it e.g.}}
\newcommand{\etal}{{\it et. al.}}
\newcommand{\ie}{{\it i.e.}}
\newcommand{\be}{\begin{equation}}
\newcommand{\dd}{\displaystyle}
\newcommand{\ee}{\end{equation}}
\newcommand{\bea}{\begin{eqnarray}}
\newcommand{\eea}{\end{eqnarray}}
\newcommand{\bef}{\begin{figure}}
\newcommand{\eef}{\end{figure}}
\newcommand{\bce}{\begin{center}}
\newcommand{\ece}{\end{center}}

\title{Thermodynamic Geometry of Strongly Interacting Matter}
\author{P. Castorina$^{1,2}$, M. Imbrosciano$^1$ and D. Lanteri$^{1,2}$}
\affiliation{
\mbox{${}^1$ Dipartimento di Fisica, Universit\`a di Catania, Via Santa Sofia 64,
I-95123 Catania, Italy.}\\
\mbox{${}^2$ INFN, Sezione di Catania, I-95123 Catania, Italy.}\\
}

\date{\today}
\begin{abstract}

The thermodynamic geometry formalism  is applied to strongly interacting matter  to estimate the deconfinement temperature. 
The curved thermodynamic metric for Quantum Chromodynamics  (QCD) is evaluated on the basis of lattice  data, whereas  the hadron resonance gas model is used for the hadronic sector.
Since the deconfinement transition is a crossover, the geometric criterion used to define the \mbox{(pseudo-)critical} temperature,  as a function of the baryonchemical potential $\mu_B$, is $R(T,\mu_B)=0$, where $R$ is the scalar curvature. The (pseudo-)critical temperature, $T_c$,  resulting from QCD thermodynamic geometry is in good agreement with lattice  and phenomenological  freeze-out temperature estimates. The crossing temperature, $T_h$, evaluated by the hadron resonance gas, which suffers of some model dependence, is larger than $T_c$ (about $20\%$) signaling remnants of confinement above the transition. 
\end{abstract}
 \maketitle

\section*{Introduction}

Quantum Chromodynamics (QCD) at high temperature and low baryon density shows a transition from hadrons to a phase of deconfined quarks and gluons, i.e. a quark-gluon plasma (QGP). The initial idea of a deconfinement first order phase transition to a state of weakly interacting quarks and gluons has been now modified and there are clear indications that, near the critical temperature, $T_c$, the system is strongly interacting and the transition is, indeed, a crossover. The estimate of the  (pseudo-)critical temperature, $T_c = 154 \pm 9$ MeV, at $\mu_B=0$, is based on the lattice results on the chiral susceptibility~\cite{lat2} 

A different phenomenological approach to get informations on  the deconfinement transition is the statistical hadronization model~(SHM)~\cite{napoco} where one evaluates the hadronization temperature by the yields of particle species in high energy collisions (where the QGP is presumably formed).  The hadronization temperature as a function of the baryonchemical potential (the freeze-out curve) in the SHM does not exactly agree with lattice results (however see Ref.~\cite{blei}).

In this paper we discuss an alternative method,  briefly introduced in ref.~\cite{CIL1}, to evaluate the deconfinement temperature based on thermodynamic geometry, which applies the formalism of Riemannian geometry to describe thermodynamic states and phase transitions. 
The curved metric for QCD thermodynamics is evaluated on the basis of lattice  data, whereas  the hadron resonance gas (HRG) model is used for the hadronic sector and
the results for the crossing temperature as a function of $\mu_B$ are in good agreement (within $10\%$) with lattice and SHM estimates. 

Thermodynamic geometry is described in details in Sec.~\ref{se:TG}, since its final (and consistent) version is fairly recent; Sec.~\ref{sec:R} contains the definition and the physical meaning of the fundamental quantity, $R$, the scalar curvature, and the discussion of the different criteria of definition of phase transitions.
Since QCD lattice simulations are reliable at small baryon density, in Sec.~\ref{sec:TR} a general scheme to evaluate $R$ as a power series in $\gamma^2=(\mu_B/T)^2$ is  introduced. Sec.~\ref{sec:QCD} and \ref{sec:HRG} are respectively devoted to the  geometric description of QCD and HRG thermodynamics. Sec.~\ref{sec:TT} contains the evaluation of the crossing temperature and in Sec.~\ref{sec:CC} some final comments are proposed. 

\vfill

\section{Thermodynamic Geometry\label{se:TG}} 

The introduction of Riemannian geometry to the analysis of thermodynamic phase space is not intuitive and the concept of distance between equilibrium configurations requires a in-depth study. Nevertheless, it  turns out to be a useful and predictive tool for thermodynamical systems.

The first application of differential geometry to statistical systems dates back to 1945 with a seminal paper by the Indian mathematician C.~Rao~\cite{Rao} who started the entire branch of information theory called ``information geometry''~\cite{Amari}, while the first metric structure  for thermodynamic systems is due to  F. Weinhold~\cite{wein1975}.

Weinhold's main idea has been to represent differentials of thermodynamic functions as elements of a vector space and then to define an inner product: the matrix elements  of the metric, $g_{ij}$,  were introduced as the second derivatives of the internal energy with respect to extensive parameters. In this formulation, the minimum energy principle for an isolated system is the basis of the geometry, implying the tensor character of $g_{ij}$ and its euclidean character. 
Despite the interesting aspects  of this approach, which permits to derive the basic laws of equilibrium thermodynamics from the geometric postulates, it didn't produce any significant result.

Some years later, shifting from the energy to the entropy representation, G. Ruppeiner~\cite{Ruppeiner1979} was able to create a thermodynamic geometry with a clear physical meaning. He defined the metric tensor as the Hessian of the entropy density and noticed that the resulting line element, i.e. the infinitesimal distance between neighboring equilibrium states, is in inverse relation with the fluctuation probability defined by the classical theory: a spontaneous fluctuation between points in phase space is less likely when they are far apart.

The previous concept of thermodynamic metric gave rise to some interesting developments in finite-time thermodynamics, where the increase in entropy due to non-equilibrium aspects can be related with the geodetic distance between the initial and final states of a real process~\cite{Salamon1983}.

Moreover, it has been shown that Weinhold's and Ruppeiner's metrics are conformal~\cite{Salamon1984} and both are limiting cases of  Rao's metric ~\cite{Crooks}.

The main result of thermodynamic geometry within Ruppeiner's formulation is the ``interaction hypothesis'' which states that the absolute value of the scalar curvature $R$, calculated by the metric, is proportional to the cube of the correlation length, $\xi^3$, of the underlying thermodynamic system. 
This liaison has been initially suggested  by the observation that the riemannian manifold of a classic ideal gas is flat, and $|R|$ calculated for a Van der Waals gas diverges at the liquid-vapor critical point exactly with the same exponent of $\xi^3$, predicted by the scaling laws.
The interaction hypothesis, which  recalls the well known connection between interaction and curvature of General Relativity,  stimulated many authors to evaluate $R$ for several statistical models with interesting results. 

\subsection{Differential Geometry and Fluctuation Theory}

The Classical Fluctuation Theory (ClFT) defines a probability distribution for the equilibrium thermodynamic states and it is based on the same principle of statistical mechanics, but from a different perspective.

Let $A_0$ be an isolated system with very large volume $V_0$ (universe) and $A$ an open subsystem of fixed volume $V$. We use the reference frame of the \textit{standard densities} $a \equiv (a^0, a^1 \dots a^r)$, where $a^0$ is the internal energy density and the other components are the number of particles of the different species.

The probability density to find $A$ in the point $a$ is given by
\begin{equation}
\label{simpleP}
P(a) d^n a = C e^{S_0(a)} d^n a \;,
\end{equation}
where $S_0$ is the total entropy of the universe formally regarded as an exact function of the parameters of $A$ and C is a normalization constant. Of course, the equilibrium configuration maximizes the value of $S_0$, but this method allows to expand classical thermodynamics giving a quantitative description of the fluctuations around an isolated equilibrium state~\cite{Landau1977, GreeneCallen1951}.

By the hypothesis of homogeneity of $A$ and $A_0$, since the entropy is additive and the standard extensive parameters (internal energy and particle numbers) are both additive and conserved, it's quite easy to show \cite{Ruppeiner:1995zz} that
\begin{equation} \label{homogeneity}
S_0(a) = V_0 s(a_0) + \frac{1}{2} V \left. \frac{\partial^2 s}{\partial a^\mu \partial a^\nu} \right|_{a_0} \Delta a^\mu \Delta a^\nu \;,
\end{equation}
where $s$ is the entropy density of the subsystem $A$ and $a_0$ is the state of the universe, which is an extremal point for $S_0$ (then the homogeneity implies that $a = a_0$ is the point of equilibrium between $A$ and $A_0$).

Let us now pay attention to general transformations of thermodynamic coordinates, $y = y(x)$,  that are continuous, differentiable and with nonzero Jacobian in the whole phase space (with the exception of special states like critical points).
Notice that the expression~\eqref{simpleP} is not covariant ($S_0$ is a state function but the volume element $d^n x$ is not invariant).
The transformation rules for the Hessian of the universe's entropy are
\begin{equation}
\label{tensor}
\frac{\partial ^2 S_0}{\partial x^\mu \partial x^\nu} = \frac{\partial y^\sigma}{\partial x^\nu} \frac{\partial y^\rho}{\partial x^\mu} \frac{\partial ^2 S_0}{\partial y^\rho \partial y^\sigma} + \frac{\partial S_0}{\partial y^\sigma} \frac{\partial ^2 y^\sigma}{\partial x^\mu \partial x^\nu}\;.
\end{equation}
If $x_0 = x(a_0)$ is an extremal point for $S_0$, due to the maximum entropy principle, first order derivatives vanish and Eq.~\eqref{tensor} becomes the transformation rule for the components of a second rank tensor.
Thus, we can define the metric tensor\footnote{It's easy to see, combining equations~\eqref{tensor} and~\eqref{homogeneity}, that the quantities defined in Eq.~\eqref{g} are the components of a metric tensor defined over the phase space of $A$.}
\be
\label{g}
g_{ij} = - \frac{1}{k_B} \left. \frac{\partial ^2 s(a)}{\partial a^i \partial a^j} \right|_{a_0}
\ee
and, if we impose that $x_0$ is a point of maximum for $S_0$, the quadratic form 
\be
(\Delta l)^2 \equiv g_{\mu\nu}(x_0) \Delta x^\mu \Delta x^\nu
\ee
defines a positive-definite Riemannian metric on the space of thermodynamic states.

The classical normalized fluctuation probability density in Gaussian approximation is given by
\begin{equation}
\begin{split}
\label{gaussiana}
P(x,x_0) d^n x &= \left( \frac{V}{2 \pi}\right) ^{n/2} \sqrt{g(x_0)} \times\\
&\!\!\!\!\!\times \exp \left[-\frac{V}{2} g_{\mu\nu} (x_0) \Delta x^\mu \Delta x^\nu \right]  d^n x\;,
\end{split}
\end{equation}
which is covariant ($\sqrt{g} \; d^nx$ is the invariant volume element of the phase space) and clarifies the meaning of ``more distant, less probable a spontaneous fluctuation between states''.

Greene and Callen \cite{GreeneCallen1951} showed that  the ClFT is completely equivalent to statistical mechanics in its full form, while in gaussian approximation the equivalence holds up to second fluctuation moments, but not at  higher orders.

\subsection{New Fluctuation Theory}

The central role of the distribution $P(x,x_0)$ for the meaning of thermodynamic distance suggested to revise the fluctuation theory~\cite{Ruppeiner:1995zz}
due to the several shortcomings (first of all the lack of covariance) of the classical theory, which inhibited a coherent geometric method. 
 
The starting point for the new theory is a Fokker-Planck like partial differential equation for the probability $P$,
\begin{equation}
\label{equaz}
\frac{\partial P}{\partial t} = -\frac{\partial}{\partial x^\mu} \left[ K^\mu (x) P \right] + \frac{1}{2} \frac{\partial ^2}{\partial x^\mu \partial x^\nu} \left[ g^{\mu \nu} (x)P \right]\;,
\end{equation}
where $t \equiv 1/V$, $K^\mu$ are  coefficients (for a complete explanation see ~\cite{Ruppeiner:1995zz}) and $g_{\mu\nu}$ is the inverse of the metric~\eqref{g} in order that the new theory reduces to the classical one in the thermodynamic limit. Notice that the tensor character of $g^{\mu\nu}$ emerges as a direct consequence of the covariance of the fluctuation equation.

In the new approach, called the Covariant and Consistent Fluctuation Theory, the absolute value of $R$ is a threshold point for the scale length of the system: if $V >> |R|$, the complete solutions of the fluctuation equation are well approximated by the classical gaussian \eqref{gaussiana}. On the other hand, one knows that the classical theory is good  in the thermodynamic limit only, i.e. when the typical correlation length in the system is much smaller than $V$. This property of $R$ supports the interaction hypothesis.

\section{The Scalar Curvature $R$\label{sec:R}}
The scalar curvature $R$ is a well-known quantity defined as the trace of the \textit{Ricci} tensor and in two dimensions  contains all the information about the geometry. For example, for the two-sphere of radius $r$ its value is $R = -2/r^2$ (in the Weinberg sign convention).
In our evaluation of $R$ we will use the standard intensive quantities in the entropy representation
\be 
F^\mu \equiv \frac{\partial s(a)}{\partial a^\mu} = \left( \frac{1}{T}, -\frac{\mu^1}{T} \dots -\frac{\mu^r}{T} \right)\;,
\ee 
where $\mu^i$ are the chemical potentials of the different species and $T$ is the overall temperature. In this ``frame'', the metric depends on the derivatives of the thermodynamic potential $\phi = P/T$, where $P$ is the total pressure of the system~\cite{Ruppeiner:1995zz}. 

In two dimensions the expression for $R$ is considerably simplified:
\be\label{eq:R}
R = \frac{k_B}{2} 
\begin{vmatrix}
	\phi_{,11} & \phi_{,12} & \phi_{,22} \\
	\phi_{,111} & \phi_{,112} & \phi_{,122} \\
	\phi_{,112} & \phi_{,122} & \phi_{,222}
\end{vmatrix} / 
\begin{vmatrix}
	\phi_{,11} & \phi_{,12} \\
	\phi_{,21} & \phi_{,22} 
\end{vmatrix} ^2  \;,
\ee
where $k_B$ is the Boltzmann's constant, 
\begin{equation}
g = \begin{vmatrix}
\phi_{,11} & \phi_{,12} \\
\phi_{,21} & \phi_{,22} 
\end{vmatrix}
\end{equation}
is the determinant of the metric and the usual comma notation for derivatives has been used ( for example $\phi_{,12}$ indicates the derivative of $\phi$ with respect to the first coordinate $\beta=1/T$ and the second coordinate ($\gamma=-\mu/T$)).  	

As already mentioned, the first confirmations~\cite{Ruppeiner1979} for the interaction hypothesis came from the study of the classic ideal gas,  represented by a flat space, and of a Van der Waals Gas, for which, near the liquid-vapor critical point, $R \sim t^{-2}$ ($t = (T-T_c)/T_c$ is the reduced temperature), exactly as expected from scaling laws if $R \propto \xi ^3 $.
Direct calculations of $R$ for other known models give reliable indications: $R$ shows a very good corrispondence with $\xi^3$ over large regions in phase space in the Takahashi Gas~\cite{Ruppeiner:1995zz} and in the ferromagnetic monodimensional Ising model~\cite{Janyszek1990b}.

This possibility to  estimate the correlation length with no, a priori, knowledge of the microscopic structure of the system is very appealing and, indeed, stimulated many applications in the contexts of pure fluids, black holes thermodynamics and critical phenomena. 

In particular, very interesting results have been obtained in the field of real fluids. The rationale is that the absolute value of $R$ is a direct measure of the size of organized mesoscopic fluctuating structures in thermodynamic systems~\cite{Ruppeiner:2017}. The integration of the phase diagram with $R$-contours ($R$-diagrams) led to the identification of several characteristic areas, corresponding to specific features of the substances, with remarkable results in the case of water~\cite{May2015}.
Moreover, the identification of $R$ with $\xi^3$ allows for a direct computation of the Widom line~\cite{Ruppeiner:2011gm} through the isobaric maxima of $|R|$, which has been explicitly evaluated for Helium, Hydrogen, Neon and Argon with good agreement with experimental data.

\subsection{R-crossing Method\label{sec:RCM}}

The correlation between $R$ and $\xi^3$ led~\cite{Ruppeiner:2011gm} to a new method to characterize the first order phase transitions. In fact,  starting from the Widom's microscopic description of the liquid-gas  coexistence region, i.e. from the idea that the correlation lengths of the two phases must be the same at the transition, one concludes that also $R$ has to vary with continuity in a first-order transition.

The previous consideration suggested an analytical method, called~\textit{$R$-Crossing}: knowing the representation of thermodynamic quantities in the two phases, one can build up the transition curve by imposing the continuity of $R$.  In other terms, the location of the coexistence curve of a first-order phase transition can be obtained from the equality of $R$ calculated in the two different phases.

In ref.~\cite{Ruppeiner:2011gm}, by using the two physical branches of the Van Der Waals model as separate representations for the liquid and vapor phase,   the value of $R$ has been evaluated along different isotherms. For a given temperature, the value of the pressure that realizes the crossing between the curvatures is selected to be the point of transition. This method has led to quantitative improvements 
with respect to the Maxwell's construction in fitting the experimental data for different real fluids.

Moreover the $R$-crossing method has been tested in systems with different features: in~\cite{Ruppeiner2013,May2012} it has been applied to construct the vapor-liquid coexistence line for the Lennard-Jones fluids, finding striking agreement with  other methods;
in~\cite{Dey:2011cs} the authors studied the geometry of the thermodynamics of first and second order phase transitions of mean-field Curie-Weiss model (ferromagnetic systems) and also of liquid-liquid phase transitions. Another field of application of the $R$-crossing method is the study of phase transitions of cosmological interest:  in~\cite{Chaturvedi:2014vpa} the authors studied the liquid-gas like fist order phase transition in dyonic charged AdS black hole and in~\cite{Sahay:2017hlq} the Hawking-Page transitions in Gauss-Bonnet-AdS black holes.

\subsection{Sign of $R$ and $R=0$ criterion}
All the cited papers in Sec.~\ref{sec:RCM} concern first order phases transitions. However, two different  phases could be related by a cross-over, as for the QCD deconfinement phenomenon, and a different  criterion,based on the sign of $R$, can be introduced to study this kind of physical behavior.

Within the thermodynamic geometry approach, the physical meaning of the sign of $R$ is still under debate but there are indications that it is directly related to the microscopic interactions.
 
More precisely, some calculations concerning pure fluids  reveal that most of the liquid and gaseous regions in the $T-P$ phase diagram, where 
$R <0$, correspond to sufficiently large average molecular separation distances where the attractive part of the intermolecular interaction potential dominates. There are, however, fluid states with positive $R$, which typically occur at large densities with repulsive intermolecular interactions.
In this context, the $R=0$ curves are able to analytically identify some anomalous behaviors observed in experimental data of several substances (in particular of water)~\cite{Ruppeiner:2017,Rupp2012b}. 

Moreover, the thermodynamic scalar curvature  for the Lennard-Jones system exhibits  a transition from $R> 0$ to $R < 0$ when the attraction in the intermolecular potential dominates~\cite{Ruppeiner2013,May2012}.

A similar behavior has been found for quantum gases, but with a different meaning:  $R$ is positive for fermi statistical interactions  and it is negative in the bosonic case~\cite{Janyszek1990}.
Anologous results apply for ideal quantum gases obeying Gentile's statistics~\cite{Oshimadag:1999} and for quantum group invariant systems~(see~\cite{Ubriaco:2016} and references therein).

\begin{figure}
	\centering 
	\includegraphics[width=0.8\columnwidth]{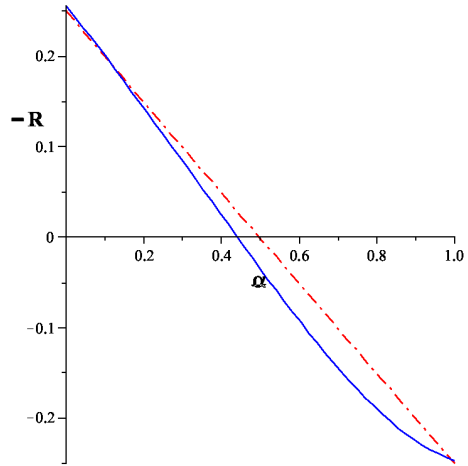}
	\caption{ $-R$ for an ideal anyon gas of particles obeying fractional statistics as a function of the parameter $\alpha$ that specifies the particle content: $\alpha=0$ corresponds to bosons, $\alpha =1$ to fermions, and $0<\alpha<1$ to intermediate statistics.  The dot-dashed line is for the classical limit and the continuous one shows the change in $R$ due to non-classical behavior. Figure from~\cite{Mirza:2008fy}, where their scalar curvature corresponds to $-R$ with our definition.}
	\label{fig:Ro}
\end{figure}

A interesting analysis concerns an anyon gas ~\cite{Mirza:2008fy} with a parametric statistical distribution given by
\begin{equation}
n_i=\frac{1}{e^{(e_i-\mu)/T}+2\,\alpha-1} \;,
\end{equation}
where $\alpha$ is the parameter that specifies the statistical behavior ($\alpha=0$ corresponds to bosons, $\alpha =1$ to fermions, and $0<\alpha<1$ to intermediate statistics). 
The sign of $R$ changes at $\alpha=1/2$ in the classical limit (dot-dashed line in Fig.~\ref{fig:Ro}) and the $R=0$ condition is satisfied by slightly lower values of $\alpha$ (continuous line) when deviations from the classical behavior are included (see ref.~\cite{Mirza:2008fy} for details). 

Finally, the sign of $R$ can hide information on the underlying interactions for black holes. 
For example, in~\cite{Sahay:2010tx} one shows that the scalar curvature remains negative for the metastable phase of the black hole, but changes sign  at the
Hawking-Page transition temperature that, therefore, can be  associated with the condition $R = 0$.

\section{Power series expansion of the scalar curvature in 2D\label{sec:TR}}

In this paper we investigate the thermodynamic geometry of the deconfinement transition by considering two thermodynamic variables, $\beta=1/T$ and $\gamma^2$, i.e.  a 2-dimensional thermodynamic metric (see Eq.~\eqref{eq:R}).

In the high temperature  regime the  phase of strong interacting matter is described by QCD lattice simulations,
reliable at low baryon density, and therefore in the calculation of the  potential $\phi=P/T$ we consider a power series expansion in $\gamma^2 < 1$.

By the expression of the pressure $P$  as a power series around the point $\mu_B=0$, 
\begin{equation}\label{eq:p}
P(\beta,\gamma) = P_0 + P_2\,\gamma^2 + 
P_4\,\gamma^4
+
P_6\,\gamma^6
+
\cdots\ ,
\end{equation}
 the thermodynamical potential $\phi=P/T$, the metric tensor $g_{ij}=\phi_{,ij}$ and the scalar curvature $R$ can be express by analogous power series (see App.~\ref{app:TR}), i.e. 
\begin{equation}\label{eq:tp}
\phi(\beta,\gamma) = A(\beta) + B(\beta) \gamma^2 + 
C(\beta) \gamma^4 + D(\beta) \gamma^6+\cdots \;,
\end{equation}
\begin{equation}\label{eq:Rseries}
R(\beta,\gamma)
=
\sum_{n=0} R_{\mathcal O(2\,n)}(\beta)
\;
\gamma^{2\,n}\;.
\end{equation}
The coefficients of the thermodynamical potential are given by
\begin{equation}\label{eq:ABCD}
\begin{split}
A(\beta)=  & P_0(\beta)\;\beta \;,
\\
B(\beta) =  & P_2(\beta)\,\beta = \frac{\chi_2(\beta)}{2!\;\beta^3} \;,
\\
C(\beta) =  & P_4(\beta)\,\beta =\frac{\chi_4(\beta)}{4!\;\beta^3} \;,
\\
D(\beta) =  & P_6(\beta)\,\beta =\frac{\chi_6(\beta)}{6!\;\beta^3} \;,
\end{split}
\end{equation}
where $\chi_{2n}=\frac{\partial^{2n}}{\partial \gamma^{2n}}(P\;\beta^4)|_{\gamma=0} = (2n)!\; P_{2n}\,\beta^4$.

The coefficients $R_{\mathcal O(2n)}$ are functions of $A$, \mbox{$B$, $\ldots$} in Eqs.~(\ref{eq:tp},~\ref{eq:ABCD}) and of their derivatives with respect to $\beta$. Particularly, one can see that the 
$2n$-coefficient $R_{\mathcal O(2n)}$ is a function of the first $2(n+1)$ coefficients of the expansion for the potential $\phi$ in  Eq.~\eqref{eq:tp}. 
For example, the zero-order term,  $R_{\mathcal O(0)}$, depends on the first and second coefficients of the $\phi$ series expansion and it is given by
\begin{equation}
\begin{split}
R_{\mathcal O(0)} = &
\frac{1}{2} 
\,
\frac{B^\prime}{A^{\prime\prime}\,B} 
\left(
\frac{A^{\prime\prime\prime}}{A^{\prime\prime}}
-
\frac{B^\prime}{B} \right)
=\\
=&
\frac{1}{2\;\ddot P_0} 
\;
\left[3 + T\;\frac{\dot \chi_2}{\chi_2}\right]
\left[
\frac{\dddot P_0}{\ddot P_0} 
- \frac{\dot \chi_2}{\chi_2} 
\right]\;,
\end{split}
\end{equation} 
where ``$\;^\prime\;$'' and ``$\;\dot \;\;$'' denote, respectively,  the derivative with respect to $\beta$ and  $T$;  $P_0(\beta)$ is the pressure and $\chi_2(\beta) = \partial^2 (P/T^4)/\partial \gamma^2$,  both at $\mu_B=0$. The other terms are evaluated in App.~\ref{app:TR}.

In conclusion, if one knows the pressure up to $\gamma^6$, $g$ and $R$ can be calculated  up to $\gamma^4$.

\section{Thermodynamic Geometry of QCD\label{sec:QCD}}

Following the results of the QCD lattice (L) simulations in ref.~\cite{Bazavov:2017dus}, the expansion series for the pressure is
\begin{equation}\label{eq:SP}
P^L(\beta,\gamma)\simeq 
P^L_0 + \sum_{n=1}^{\infty} \frac{P^L_{2n}}{\beta^4}\,\gamma^{2n} 
\end{equation}
and, by comparison with eq.(13), one gets
\begin{equation}
\begin{split}
A^L=&P_0^{L}\;\beta \;,\\
B^L=&\frac{P_2^{L}}{\beta^3} \;,\\
C^L =&\frac{P_4^{L}}{\beta^3}\;,  \\
D^L =&\frac{P_6^{L}}{\beta^3}  \;,
\end{split}
\end{equation}
where, for strangeness neutral systems with a fixed ratio of electric charge to baryon density (see~\cite{Bazavov:2017dus} for details), one has
\begin{equation}
P_2^L(\beta)=\frac{1}{2}\left[N_1^B(\beta) + r\;q_1(\beta)\,N_1^B(\beta)\right] 
\end{equation}
\begin{equation}
\begin{split}
P_4^L(\beta)=\frac{1}{4}\Big[&N_3^B(\beta) + r\;\Big(q_1(\beta)\,N_3^B(\beta) +\\
&\qquad
 + 3\,q_3(\beta)N_1^B(\beta)\Big)\Big] 
\end{split}
\end{equation}
\begin{equation}
\begin{split}
P_6^L(\beta)=\frac{1}{6}\Big[&N_5^B(\beta) + r\;\Big(q_1(\beta)\,N_5^B(\beta) +\\
&
+ 3\,q_3(\beta)N_3^B(\beta+5\;q_5(\beta)\,N_1^B(\beta))\Big)\Big] \;,
\end{split}
\end{equation}
being $N_{2n-1}^B$ the $(2n-1)$-th coefficient for the power expansion of the baryon number density divided by $T^3$, 
\begin{equation}
\frac{n_B}{T^3}
=
\sum_{n=1}^{\infty}
N_{2n-1}^B\;\gamma^{2n-1}\;,
\end{equation}
 $q_{2n-1}$ are the expansion coefficients for the electric charge chemical potential, and 
\begin{equation}
r \equiv \frac{n_Q}{n_B}\;,
\end{equation}
with $n_Q$ and $n_B$ the charge and baryon number densities respectively. 

Three special cases are considered: the electric neutral systems, $r=0$, the isospin symmetric limit $r=1/2$, i.e. $q_k=0\;\forall k$, 
which gives the same result of $r=0$, and $r=0.4$, usually considered for applications to heavy ion collisions~\cite{Bazavov:2017dus,Bazavov:2012vg}.

\begin{figure}
	\centering 
	\includegraphics[width=\columnwidth]{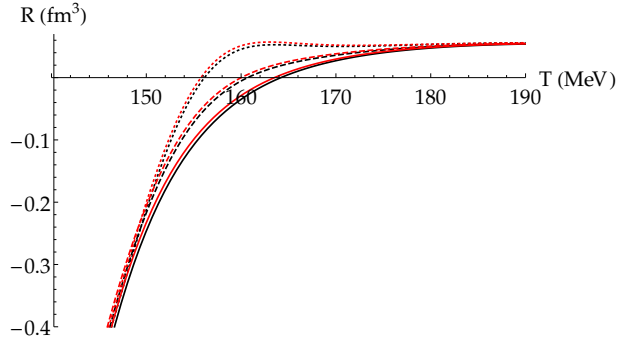}
	\caption{The scalar curvature $R$ from Eq.~\eqref{eq:Rseries}: 
		the black curves are for lattice data obtained for the condition $n_S=n_Q=0$ (or equivalent for the isospin symmetric limit), while the reds are for $n_S=0$ and $n_Q/n_B =0.4$ ~\cite{Bazavov:2017dus,qm18}.
		The continuous lines are for $\mu_B=0$ MeV, the dashed ones for $\mu_B=80$ MeV and the dotted lines for $\mu_B=135$ MeV. }
	\label{fig:R}
\end{figure}

Figure~\ref{fig:R} shows the scalar curvature $R$ evaluated by Eq.~\eqref{eq:Rseries} and Eqs.(17-23). The black curves are based on lattice data with the condition $n_S=n_Q=0$ (or equivalent for the isospin symmetric limit), whereas the red ones are for $n_S=0$ and $n_Q/n_B =0.4$.
The continuous lines are for $\mu_B=0$ MeV, the dashed ones for $\mu_B=80$ MeV and the dotted lines for $\mu_B=135$ MeV. 

\section{Thermodynamic Geometry of the Hadron Resonance Gas\label{sec:HRG}} 

The confined phase can be described in terms of a non-interacting gas of hadrons. There are several versions of the HRG which give different results~\cite{ambi} with some ambiguity and dependence on the specific model.

In the HRG model with point-like constituents, if $m_{max}$ is the maximum mass one includes,  the trace anomaly can be written as a sum over all particles species with mass $m_i\leq m_{max}$~\cite{Karsch:2003vd}
\begin{equation}\label{eq:HRG0}
\begin{split}
\left(\frac{\Theta^{\mu\mu}}{T^4}\right)^{H}
&=
\sum_{m_i\leq m_{max}}\frac{d_i}{2\,\pi^2} \times
\\
& \times \sum_{k=1}^{\infty} 
\frac{(-\eta_i)^{k+1}}{k}\left(\frac{m_i}{T}\right)^3K_1\left(\frac{k\,m_i}{T}\right)\;,
\end{split}
\end{equation}
where $\eta_i=-1(+1)$ for bosons (fermions), $K_1$ is the modified Bessel function, $d_i$ are the degeneracy factors.

For small $\mu_B$, the baryon sector of a HRG can be described by the Boltzmann approximation and the pressure can be written as~\cite{Bazavov:2017dus}
\begin{equation}\label{eq:HRG}
P^{H}(\beta,\gamma) = P^{H}_0(\beta)
+
P_B^{H}(\beta)\left(\cosh\gamma -1 \right)\;,
\end{equation}
where $P^{H}_0(\beta)=P_M^{H}(\beta)+P_B^{H}(\beta)$ is the total pressure at $\mu_B=0$ (Eq.~\eqref{eq:HRG0}), $P_M^{H}(\beta)$ and $P_B^{H}(\beta)$ are the meson and baryon contributions to Eq.~\eqref{eq:HRG0}, respectively.

\begin{figure}
	\centering 
	\includegraphics[width=\columnwidth]{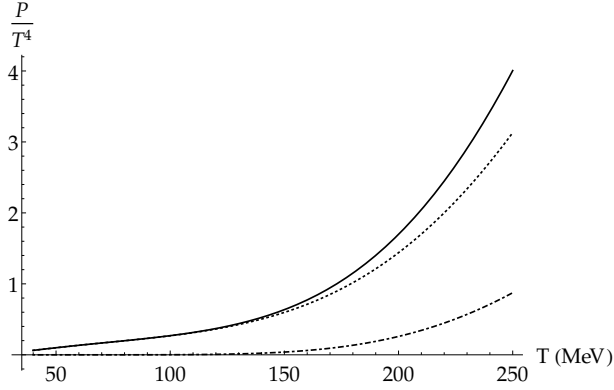}
	\caption{The total pressure of a HRG gas $P^{H}_0(\beta)$ (black line), the meson contribution $P_M^{H}(\beta)$ (dotted line) and the $P_B^{H}(\beta)$ (dot-dashed line) at $\mu_B=0$.}
	\label{fig:PHRGc}
\end{figure}

In Figure~\ref{fig:PHRGc} are plotted the total pressure $P^{H}_0(\beta)$ at $\mu_B=0$, the meson part, $P^{H}_M(\beta)$, and the baryonic part, $P^{H}_B(\beta)$.

\begin{figure}
	\centering 
	\includegraphics[width=\columnwidth]{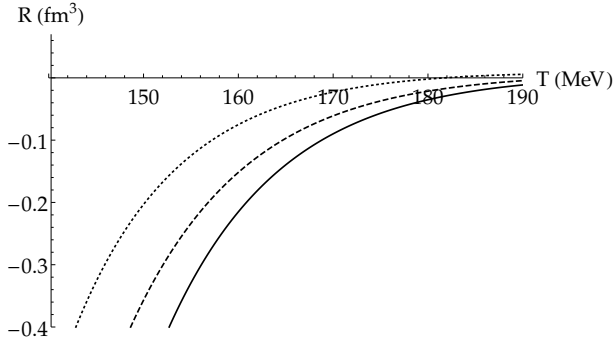}
	\caption{The scalar curvature $R$, evaluated by Eqs.~(\ref{eq:SHRG},~\ref{eq:AH}) for different values of the baryonchemical potential, $\mu_B=0$ MeV (continuous lines), $\mu_B=80$ MeV (dotted lines) and $\mu_B=135$ MeV (dashed lines), obtained by the expansion of Eq.~\eqref{eq:Rseries} at the 4-th order.}
	\label{fig:RHe}
\end{figure}

For comparison with the QCD calculations in Sec.~\ref{sec:QCD}, one evaluates the series expansion in $\gamma^2$ of Eq.~\eqref{eq:HRG} in the Boltzmann approximation (i.e. all baryon number susceptibilities are identical, $\chi^{H}_{2k}=\chi^{H}_2=2\;P_B^H\,\beta^4$) to obtain
\begin{equation}\label{eq:SHRG}
P^{H}(\beta,\gamma) \simeq P^{H}_0(\beta)
+
\frac{\chi^{H}_2(\beta)}{\beta^4}
\sum_{n=1}^{\infty}\frac{\gamma^{2n}}{(2n)!} 
\end{equation}
and the coefficients of the thermodynamical potential for the hadronic (H) sector are given by
\begin{equation}\label{eq:AH}
\begin{split}
A^H=&P_0^{H}\;\beta \;,\\
B^H=&\frac{\chi_2^{H}}{\beta^3} \;,\\
C^H =&\frac{2\;B^H}{4!} \;,  \\
D^H =&\frac{2\;B^H}{6!}  \;.
\end{split}
\end{equation}
In Fig.~\ref{fig:RHe} is plotted the scalar curvature $R$ for different values of the baryonchemical potential: $\mu_B=0$ MeV (continuous lines), $\mu_B=80$ MeV (dotted lines) and $\mu_B=135$ MeV (dashed lines), obtained by the expansion of Eq.~\eqref{eq:Rseries} at order
 $\gamma^4$.

\section{Transition temperature \label{sec:TT}}

In the QGP phase the system is mostly of fermionic type while in the confined phase is essentially a bosonic (mesonic) one. Moreover one knows from lattice simulation that the transition is a cross-over and therefore, following the previous discussion, the crossing  temperature from QGP to a confined mesonic system can be evaluated by implementing the condition  $R=0$. 

\begin{figure}
	\centering 
	\includegraphics[width=\columnwidth]{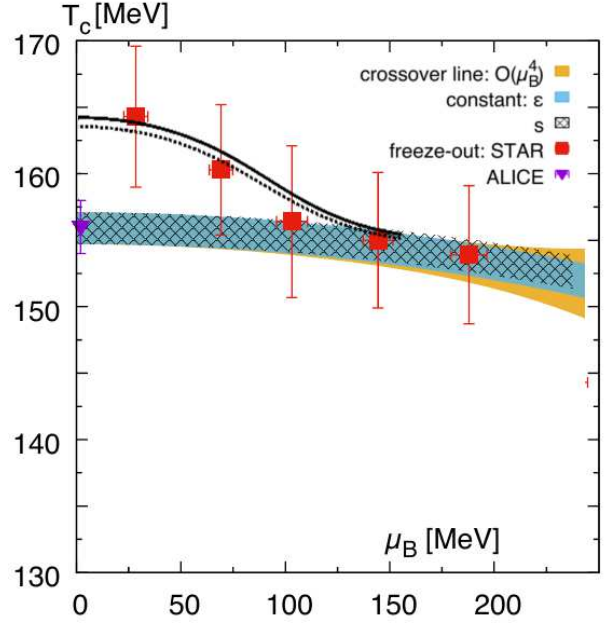}
	\caption{The crossing temperature, both for $r=0$ or $r=0.5$ (continuous black line) and for $n_S=0$ and $n_Q/n_B=0.4$ (black dotted line), compared with lattice data (see text~\cite{qm18})  and the results of the freeze out temperature from ALICE (purple point~\cite{Floris:2014pta}) and STAR (red points~\cite{Das:2014qca,Adamczyk:2017iwn}) collaborations.}

	\label{fig:TR}
\end{figure}

In Figure~\ref{fig:TR} is plotted the critical temperature $T_c(\mu_B)$ at which the scalar curvature of the QGP phase  crosses the $R=0$ line, both for $r=0$ or $r=0.5$ (continuous black line) and for $n_S=0$ and $n_Q/n_B=0.4$ (black dotted line), compared with lattice results~\cite{Bazavov:2017dus,qm18} and the freeze-out temperature  obtained by ALICE~\cite{Floris:2014pta} and STAR~\cite{Das:2014qca,Adamczyk:2017iwn} collaborations. The yellow curve gives the crossover temperature, the blue and black grid bands are obtained by considering fixed values of the energy density (blue) or of the entropy density (grid) (see Ref.~\cite{Bazavov:2017dus,qm18} for details).  

The same criterion, $R=0$ applied to the HRG gives the crossing from a mostly mesonic system to a fermion dominated one at temperature, $T_h$, about 20$\%$ larger than $T_c$. 

The two different temperatures have a possible interpretation if one recalls that, since the deconfinement transition is a cross-over, one can expect remnants of confinement slightly above $T_c$. Indeed the persistence of string-like objects above $T_c$ has ben obtained by many different methods: lattice simulations \cite{karsch,cea}, quasiparticle approach \cite{mannarelli1,mannarelli2}, NJL correlator \cite{beppe,jap}, Mott transitions \cite{david} and confinement mechanisms \cite{eddy}. 

Following this interpretation, $T_c$ is the deconfinement temperature and $T_h$ is the temperature of the complete melting of a light meson.

\section{Comments and Conclusions\label{sec:CC}}

Thermodynamic geometry applied to QCD deconfinemt transition is a useful tool to evaluate the transition temperature $T_c$. The results, obtained by the criterion $R=0$, are in good agreement with lattice data and freeze-out calculations in the low density region. However the criterion is completely general and can be applied at large baryon density if the potential $\Phi$ can be estimated in a reliable way.
On the other hand the  temperature, $T_h$, evaluated by HRG is larger than $T_c$, suggesting the interpretation that the meson melting temperature is larger than the  temperature associated with the chiral susceptibility. However this conclusion could be model dependent, because we have used a specific model of the HRG. The introduction of other dynamical details, as the excluded volume, could change the HRG evaluation of $R$, by including some effective repulsive interaction similar to Fermi statistic effects and then closing, in part, the gap with the value of $R$ for a fermionic system. This analysis will be carried out in a forthcoming paper.  

\section{Acknowledges}

The authors thank F.Karsch for the HRG data (in Fig.~\ref{fig:PHRGc}) and H.Satz for useful comments. P.Castorina thanks the Theoretical High Energy Physics Group of Bielefeld University for the hospitality.

\appendix
\section{Power series of the scalar $R$\label{app:TR}} 
Let us consider the power series expansions in $\gamma^2$ for the pressure, i.e. 
\begin{equation}\label{eq:sp}
\begin{split}
P(\beta,\gamma)
=&
P_0(\beta)
+
P_2(\beta)\,\gamma^2
+
P_4(\beta)\,\gamma^4
+\\
&+
P_6(\beta)\,\gamma^6
+
P_8(\beta)\,\gamma^8
+\cdots \;,
\end{split}
\end{equation}
and for the potential $\Phi$,
\begin{equation}
\begin{split}
\phi(\beta,\gamma)
=&
A(\beta)
+B(\beta)\,\gamma^2
+C(\beta)\,\gamma^4
+\\
&+D(\beta)\,\gamma^6
+E(\beta)\,\gamma^8
+\cdots\;.
\end{split}
\end{equation}
The metric element $g_{11}=\phi_{11}$ and the determinant  $g$ can be written respectively as
\begin{equation}
\begin{split}
g_{11}(\beta,\gamma)
=&
A^{\prime\prime}
+B^{\prime\prime}\,\gamma^2
+
+C^{\prime\prime}\,\gamma^4
+\\
&
+D^{\prime\prime}\,\gamma^6
+\cdots\;,
\end{split}
\end{equation}
\begin{equation}\label{eq:sg}
\begin{split}
g(\beta,\gamma)
=&
g_{\mathcal O(0)}
+
g_{\mathcal O(2)}\,\gamma^2
+
g_{\mathcal O(4)}\,\gamma^4
+\\
&
+
g_{\mathcal O(6)}\,\gamma^6
+\cdots\;,
\end{split}
\end{equation}
where
\begin{equation}
\begin{split}
g_{\mathcal O(0)}(\beta)
=&
2 B\;A''\;,
\end{split}
\end{equation}
\begin{equation}
\begin{split}
g_{\mathcal O(2)}
=&
2 \left(6 \;C\,A'' + B\,B'' - 2\;B'^2\right)\;,
\end{split}
\end{equation}
\begin{equation}
\begin{split}
g_{\mathcal O(4)}
=
2 \Big(& 15\,D\,A'' + 6\,C\,B'' - \\
&\quad 
 -8\,B'\,C' 
 + B\,C''\Big)\;,
\end{split}
\end{equation}
\begin{equation}
\begin{split}
g_{\mathcal O(6)}
=&
2 \Big(28\,E\,A'' + 15\,D\,B'' - 12\,B'\,D' +\\
&\qquad
+ B\,D'' + 6\,C\,C'' - 8\,C'^2\Big)
\end{split}
\end{equation}
and the symbol ``$^\prime$'' indicates the derivative with respect to $\beta$.

From previous equations it turns out that the $2n$-th term of the series expansion of Eq.~\eqref{eq:sg} depends on the first $2(n+1)$-th terms of the expansion of $P$. Therefore, if we know the pressure up to  $\gamma^6$, $g$ and $R$ can be evaluated at most up to  $\gamma^4$.

To show that Eq.~\eqref{eq:sp} leads to a similar series expansion for $R$,  i.e. 
\begin{equation}\label{eq:sr}
\begin{split}
R(\beta,\gamma)
=&
R_{\mathcal O(0)}(\beta)+
R_{\mathcal O(2)}(\beta)\,\gamma^2 
+\\
& +
R_{\mathcal O(4)}(\beta)\,\gamma^4 
+
\cdots\;,
\end{split}
\end{equation}
let us define the auxiliary variable $\Gamma\equiv \gamma^2$  and consider the metric determinant and $R$  as a function o $\Gamma$, that is  
\begin{equation}
\begin{split}
g =&
\left|
\begin{array}{cc}
\phi_{11} & 2\,\sqrt{\Gamma}\;\phi_{1\widetilde{2}}\\
2\,\sqrt{\Gamma}\;\phi_{1\widetilde{2}} & 2\,\phi_{\widetilde{2}} + 4\,\Gamma\;\phi_{\widetilde{2}\widetilde{2}}\end{array}
\right|=\\
= &
2\,\phi_{11}\,\phi_{\widetilde{2}}
+
4\,\Gamma\,
\left[
\phi_{11}\,\phi_{\widetilde{2}\widetilde{2}}
- \left(\phi_{1\widetilde{2}}\right)^2\right]
\end{split}
\end{equation}
and 
\begin{widetext}
\begin{equation*}
\begin{split}
2\,g^2\,R
=&
4\,\phi_{1\widetilde{2}}\,\left( \phi_{\widetilde{2}}\,\phi_{111}
-
\phi_{11}\,\phi_{1\widetilde{2}}\right) 
+\\
& +
8\;\Gamma\,\Big(
\phi_{\widetilde{2}}\,\phi_{111}\,\phi_{1\widetilde{2}\widetilde{2}}
-
2\,\phi_{1\widetilde{2}}\,\phi_{\widetilde{2}\widetilde{2}}\,\phi_{111}
+
3\,\phi_{11}\,\phi_{\widetilde{2}\widetilde{2}}\,\phi_{11\widetilde{2}} 
+
\phi^2_{1\widetilde{2}}\,\phi_{11\widetilde{2}} 
-
\phi_{\widetilde{2}} \,\phi^2_{11\widetilde{2}}
-
2\,\phi_{11}\,\phi_{1\widetilde{2}}\,\phi_{1\widetilde{2}\widetilde{2}} 
\Big) 
+ \\
& + 16\;\Gamma^2\	\Big(\phi_{11}\,\phi_{11\widetilde{2}}\,\phi_{\widetilde{2}\widetilde{2}\widetilde{2}}+
\phi_{1\widetilde{2}}\,\phi_{11\widetilde{2}}\,\phi_{1\widetilde{2}\widetilde{2}}
+
\phi_{\widetilde{2}\widetilde{2}}\,\phi_{111}\,\phi_{1\widetilde{2}\widetilde{2}}
-
\phi_{\widetilde{2}\widetilde{2}}\,\phi^2_{11\widetilde{2}}
-
\phi_{11}\,\phi^2_{1\widetilde{2}\widetilde{2}}
-
\phi_{1\widetilde{2}}\,\phi_{111} \,\phi_{\widetilde{2}\widetilde{2}\widetilde{2}}
\Big)
\end{split}
\end{equation*}
\end{widetext}
where the subscripts ``$1$'',``$2$'' and ``$\widetilde{2}$'' indicate respectively  the derivative with respect to $\beta$,  $\gamma$ and  $\Gamma$.
This expression is formally exact and the replacement of $\Phi$ as a power series of $\Gamma$ gives the final result.

Finally the terms in Eq.~\eqref{eq:sr} are
\begin{equation}
R_{\mathcal O(0)}(\beta)=\frac{B'}{2 B A''} 
\left(\frac{A'''}{A''}
-
\frac{B'}{B}
\right)\;,
\end{equation}
\begin{widetext} 
\begin{equation}
\begin{split}
R_{\mathcal O(2)}(\beta)
=&
-\frac{B'}{2\,B A''}\left(\frac{A'''}{A''}-\frac{B'}{B}\right) \left(\frac{6 C}{B}+\frac{B''}{A''} - \frac{2 \,B'^2}{B A''}\right) 
-
\frac{B''^2-3 A''' C'}{B\,A''^2}
+
\frac{6\;C\,A''\,B''}{B^2\,A''^2}+\\
&+
\frac{B'}{2 B A''} \left(\frac{B'''}{A''}-\frac{6\,C\, A'''}{B\,A''}-\frac{12\, C'}{B}\right)
+
\frac{B'^2 B''}{2 B^2 A''^2} \;,
\end{split}
\end{equation}

\begin{equation}
\begin{split}
R_{\mathcal O(4)}(\beta)
=&
\frac{B'}{2\,B\,A''} \left(\frac{A'''}{A''}-\frac{B'}{B}\right)
\Bigg[
3 \left(\frac{6 C}{B}+\frac{B''}{A''}- \frac{2\,B'^2}{B \,A''}\right)^2
-
\frac{30\,D}{B} 
-
 \frac{12\,C\,B''}{B\, A''}
+
\frac{16\,B'\, C'}{B\, A''} - \frac{2\,C''}{A''}\Bigg]
-\\
&-
\frac{1}{B\,A''} \left(\frac{6\,C}{B} + \frac{B''}{A''}-\frac{2\,B'^2}{B\,A''}\right)
\Bigg[-\frac{2\,\left(B''^2-3 A''' C'\right)}{A''}
+ 
\frac{12\,C\,B''}{B}
+
\frac{B'^2 B''}{B\,A''}
-\\
&\qquad\quad\qquad\quad\qquad\quad\qquad\quad\qquad\quad
- 
B' \left(\frac{6 C A'''}{B\,A''}+\frac{12\,C'}{B} - \frac{B'''}{A''}\right)
\Bigg]
+\\
&+
\frac{1}{2\,B\,A''} \Bigg[\frac{15\, A''' D'}{A''} + \frac{60\,D\,B''}{ B}  
+
\frac{24\, C\, C''}{ B}-\frac{36\, C'^2}{ B}
-
\frac{45 D A''' B'}{ B A''} - \frac{30\,D'B'}{ B} 
-\\
&\qquad\quad\qquad\quad
- \frac{6\,B' C\, B'''}{ B A''} 
+
\frac{B' C'''}{A''}
-
\frac{8\, B'' C''}{ A''} + \frac{3\,B'^2 C''}{ B A''}\Bigg]+\\
&+
\frac{C'\left(6 C A'''+3 B B'''+2 B' B''\right)}{B^2 A''^2}  
\end{split}
\end{equation}

\end{widetext}

\end{document}